\documentclass[sigconf,screen]{acmart}
\usepackage{tikz}
\usetikzlibrary[graphs]

\usepackage{color}

\definecolor{pblue}{rgb}{0.13,0.13,1}
\definecolor{pgreen}{rgb}{0,0.5,0}
\definecolor{pred}{rgb}{0.9,0,0}
\definecolor{pgrey}{rgb}{0.46,0.45,0.48}

\usepackage{relsize}
\usepackage{listings}
\newcommand{\code}[1]{\lstinline~#1~}

\usepackage{amsmath}
\usepackage{url}

\usepackage{flushend}
\newcommand{\UBER}{Uber}
\newcommand{\UBERFULL}{Uber Technologies, Inc.}
\newcommand{\UBEREATS}{Uber Eats}
\newcommand{\UBERDRIVER}{Uber Driver}

\newcommand{\customSectionSpacing}{-0.3em}
\setcopyright{acmlicensed}
\acmPrice{15.00}
\acmDOI{10.1145/3238147.3238174}
\acmYear{2018}
\copyrightyear{2018}
\acmISBN{978-1-4503-5937-5/18/09}
\acmConference[ASE '18]{Proceedings of the 2018 33rd ACM/IEEE International Conference on Automated Software Engineering}{September 3--7, 2018}{Montpellier, France}
\acmBooktitle{Proceedings of the 2018 33rd ACM/IEEE International Conference on Automated Software Engineering (ASE '18), September 3--7, 2018, Montpellier, France}

\begin{document}

\ccsdesc[500]{Software and its engineering~Software notations and tools}
\ccsdesc[500]{Software and its engineering~Formal software verification}

\keywords{stream-based programming, refinement types, mobile applications}

%
\title{Safe Stream-Based Programming with Refinement Types}

\author{Benno Stein}
\affiliation{\institution{University of Colorado Boulder}\country{Boulder, Colorado, USA}}
\email{benno.stein@colorado.edu}
\author{Lazaro Clapp}
\affiliation{\institution{Uber Technologies, Inc.}\country{San Francisco, California, USA}}
\email{lazaro@uber.com}

\author{Manu Sridharan}
\affiliation{\institution{Uber Technologies, Inc.}\country{San Francisco, California, USA}}
\email{msridhar@uber.com}

\author{Bor-Yuh Evan Chang}
\affiliation{\institution{University of Colorado Boulder}\country{Boulder, Colorado, USA}}
\email{evan.chang@colorado.edu}

\begin{abstract}  
In stream-based programming, data sources are abstracted as a stream of values that can be manipulated via callback functions.
Stream-based programming is exploding in popularity, as it provides a powerful and expressive paradigm for handling asynchronous data sources in interactive software.
However, high-level stream abstractions can also make it difficult for developers to reason about control- and data-flow relationships in their programs.
This is particularly impactful when asynchronous stream-based code interacts with thread-limited features such as UI frameworks that restrict UI access to a single thread, since the threading behavior of streaming constructs is often non-intuitive and insufficiently documented.

In this paper, we present a type-based approach that can statically prove the thread-safety of UI accesses in stream-based software.
Our key insight is that the fluent APIs of stream-processing frameworks enable the tracking of threads via type-refinement, making it possible to reason automatically about what thread a piece of code runs on -- a difficult problem in general.
  
We implement the system as an annotation-based Java typechecker for Android programs built upon the popular ReactiveX framework and evaluate its efficacy by annotating and analyzing 8 open-source apps, where we find 33 instances of unsafe UI access while incurring an annotation burden of only one annotation per 186 source lines of code.
We also report on our experience applying the typechecker to two much larger apps from the \UBERFULL~codebase, where it currently runs on every code change and blocks changes that introduce potential threading bugs.\vspace{-0.5em}
\end{abstract}

\maketitle

\vspace{\customSectionSpacing}\section{Introduction}

Many popular user interface frameworks (e.g., Swing, Cocoa, Eclipse, iOS, Android) distinguish a single main thread from which all UI accesses must be performed~\cite{swing,cocoa,eclipse,ios,android_dev_guide}.
This design is preferred by library developers since it eliminates the need for library-internal synchronization: there is no need to worry about data races or deadlock when only one thread is allowed to perform UI operations.

However, the single UI thread model requires application developers to carefully avoid interacting with the UI from other threads since doing so results in a runtime crash or undefined behavior.

Such {\em invalid thread access} bugs are very common in practice: a Google search for Android's \code{CalledFromWrongThreadException}, one of several exceptions that Android can throw when the UI is accessed improperly, yields over 47,000 results, including numerous Github bug reports, StackOverflow questions, and developer guides and tutorials.

Furthermore, invalid thread accesses are difficult to detect and debug in practice.
Existing UI testing techniques are often unable to achieve adequate coverage of possible UI interaction traces \cite{DBLP:conf/kbse/ChoudharyGO15} and struggle with bugs that only manifest on certain devices in the diverse Android hardware and software ecosystem \cite{DBLP:conf/kbse/WeiLC16, DBLP:conf/kbse/FazziniO17}.
Program analysis approaches to finding improperly threaded UI accesses are similarly inadequate:
the callgraph-reachability technique proposed by Zhang et al.~\cite{DBLP:conf/issta/ZhangLE12} and the effect type system of Gordon et al.~\cite{DBLP:conf/ecoop/GordonDEG13} both identify methods that interact with the UI effectively but use a very conservative and restrictive model to determine when those methods run on the UI thread;
on the other hand, general approaches to concurrency analysis typically focus on shared memory access rather than determining the thread on which a given piece of code will run~\cite{DBLP:conf/pldi/NaikAW06, DBLP:conf/pldi/FlanaganF09}.


In recent years, there has been an explosion in popularity of stream-based programming frameworks like Reactive Extensions \cite{rx_website}, especially for interactive software like Android applications that need to respond to user input in real time.
Such frameworks provide expressive and convenient threading abstractions for manipulating streams of data and computation, but offer little in the way of tool support to help developers avoid invalid UI access by reasoning about what thread a stream is running on.

We propose in this paper a refinement type-based static analysis that identifies possible invalid thread UI accesses in stream-based Android applications, combining an effect type system that tracks the possible UI interactions performed by methods with a thread type system that encodes the possible threading behaviors of asynchronous data streams.

Our approach not only detects (or proves the absence of) improper UI access, but also helps developers understand and reason about UI interactions in their code.
Furthermore, our annotation-based approach makes that reasoning explicit and self-documenting, allowing future contributors to more easily understand, modify, or extend previously annotated code.

Our typechecker currently runs on every commit made to two major Android applications at \UBERFULL~(\UBER), blocking any changes that may introduce invalid thread access bugs.
In practice, we have found that the typechecker catches potential bugs earlier in the development process and more efficiently than existing testing and manual code review are able to.

We build on the insights of Gordon et al.~\cite{DBLP:conf/ecoop/GordonDEG13}, who first showed the applicability of effect-typing to modern UI frameworks with a distinguished UI thread, by extending their work in two key dimensions.
First, we introduce a thread type system for determining the thread on which stream operators execute.  Combining the thread type system with the effect typing of Gordon et al. enables verifying UI effect safety for a wider class of threading constructs, with almost no additional annotation burden for developers.
Second, we develop an effect inference technique for callbacks and lambda abstractions that further lowers the annotation burden of UI effect typing and improves code readability.

{\bf Contributions: } The primary contributions of our work are as follows:

\begin{itemize}\item
  We introduce a novel refinement-type system that soundly verifies that stream-processing code only accesses the UI from streams running on the Android main thread.
  We demonstrate its efficacy by implementing a static annotation-based type checker for Android applications built upon the ReactiveX Java stream-based programming framework.
    Our system statically refines the types of callbacks and lambda abstractions by the {\em effects} they may incur and data/event streams by the {\em threads} on which they may run, then verifies that non-UI thread streams never call methods with UI effect.
\item
  We analyze a corpus of $8$ open-source Android applications, as well as two large closed-source applications developed at \UBER.
  In doing so, we find that {\em (a)} improper UI access is detectable by our tool and prevalent in both open-source and closed-source codebases and {\em (b)} annotation burden on the programmer is low enough for the tool to be incorporated into a production developer workflow.
  In total, we find $33$ instances of UI-effectful callbacks running on non-UI-threaded streams in the open-source corpus.  At \UBER\ our tool runs on every code change as part of continuous integration, blocking any change that may introduce stream-based threading bugs.
\end{itemize}

\vspace{\customSectionSpacing}\section{Overview}
In this section, we provide background information about stream-based programming frameworks, the Android UI model, and refinement typechecking by applying our tool to a simple example from a user's perspective.

\vspace{\customSectionSpacing}\subsection{Reactive Extensions}
Reactive Extensions (ReactiveX)~\cite{rx_website} is a multi-language framework for asynchronous stream-based programming which allows developers to easily write code that operates over streams of events or data, composing and transforming them with various functional operators and subscribing callbacks to perform computations in response to events.

Stream processing frameworks have gained popularity in recent years due to their ability to provide a uniform interface to multiple asynchronous input sources, allowing developers to build responsive interactive applications and easily interoperate (via the stream API) with new frameworks and technologies.
Stream-based programming also encourages a functional programming paradigm, preferring composable modular computations to imperative procedures over mutable state.

A typical use-case of ReactiveX is to receive or generate an \code{Observable} stream, perform a number of operations that modify its data or thread, and then subscribe an \code{Observer} (or other callback-like object) to asynchronously consume the resulting event stream.
This so-called ``fluent'' interface, in which multiple calls are chained together, is a hallmark of the stream-based programming paradigm, combining ideas from the \code{Observer} pattern, the \code{Iterator} pattern, and functional programming~\cite{rx_website}.

\begin{sloppypar}
Take, for example, the code snippet in Fig.~\ref{fig:motivating_example}, which updates car locations on a map using the ReactiveX framework.  The \code{carLocationData} stream represents location data for some set of cars, updated periodically by a remote server.  The \code{filter} operator filters the stream to include only cars currently without a passenger.  Then, \code{observeOn} moves subsequent operations to the main thread, a requirement for performing UI updates on Android.  The \code{delay} operator introduces a delay before each location update, to allow for other processing to complete.  Finally, the anonymous function passed to \code{subscribe} invokes UI APIs to display the cars in the map.
\end{sloppypar}

A key feature of the ReactiveX fluent interface is that each operation in a chain of calls returns a new \code{Observable} instance rather than performing side effects on the receiver of the call.
This side-effect free nature of ReactiveX enables a type-based analysis, since each intermediate \code{Observable} instance in the call chain can be given a single static type that is not subject to change later in the chain.
Note that ReactiveX's API is distinct from the Builder pattern~\cite{Gamma:1995:DPE:186897}, which supports a similar call-chaining syntax but does perform side effects with each call, passing the same Builder instance through the chain.

\begin{figure}
\begin{lstlisting}
   Observable<...> carLocationData = ... ;
   carLocationData
    .filter(car -> /*car has no passenger*/)
    .observeOn(AndroidSchedulers.mainThread())
    .delay(100, TimeUnit.MILLISECONDS)
    .subscribe(
      car -> { /* display car on map */ },
      err -> { /* render error message */ })
\end{lstlisting}
\caption{Simple example usage of ReactiveX \code{Observable} streams.  Contains the distillation of a threading bug that was detected by our tool and fixed, as explained in section~\ref{sec:overview_types}.}
\label{fig:motivating_example}
\centering
\end{figure}
Though Fig.~\ref{fig:motivating_example}'s example makes use of only a few simple stream operators, ReactiveX provides a wide range, from functional programming standards like
\code{filter} (emit only those data that satisfy a given predicate)
to more exotic combinators like
\code{switchMap} (map incoming data to new streams, emitting events only from the most recent datum's stream).
Streams (i.e. \code{Observable}s) possess 340 such operators in ReactiveX Java version 2.1.12, in addition to any custom operators defined by third party libraries or more specialized types of streams.

With the expressivity of such a large and complex framework inevitably comes a steep learning curve, since it is difficult for developers to become familiar with the API.
This contributes to the preponderance of threading bugs in real-world applications using ReactiveX.
In our experiments, we find that many programs contain latent threading bugs that require a nuanced understanding of the framework to detect.

 The program in Fig.~\ref{fig:motivating_example}, for example, looks safe from improper UI access at first glance: its author was careful to \code{observeOn} the Android main thread before subscribing a callback that renders UI elements and error messages.  However, due to the threading behavior of the \code{delay} operator, the program actually accesses the UI from a background thread.

\vspace{\customSectionSpacing}\subsection{Refinement Typechecking}\label{sec:overview_types}

In spite of the intractability of determining what thread a piece of code will run on in general, stream-based frameworks like Reactive Extensions are amenable to thread analysis by means of {\em refinement types}~\cite{DBLP:conf/pldi/FreemanP91,DBLP:conf/issta/PapiACPE08}.
As discussed in the previous section, the fluent functional interface of ReactiveX streams is the key feature which enables the use of a type-based approach to track the thread of intermediate calls in the chain, obviating the need for more expensive general-purpose thread analyses.

Informally, a refinement type system can be thought of as augmenting {\em base types} (e.g., integers, lists, strings) by {\em qualifiers} that further restrict the values of the base type (e.g., {\em positive} integers, {\em nonempty} lists, {\em ASCII} strings).
Refinement subtyping holds when both base subtyping and qualifier subtyping hold, and standard intuitions about nominal subtyping over base types behave analogously for refined types%
\footnote{Namely, refinement subtyping is reflexive and transitive, and the Liskov substitution principle holds for refined types.}.

In this paper, we verify the safety of UI access in stream-based Android programs using two separate but parallel type refinements.

\begin{figure}
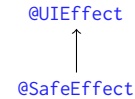

  \centering
  \tikz {
    \node (t) at (2,2) [] {\code{@UIEffect}};
    \node (b) at (2,1) [] {\code{@SafeEffect}};
    \draw (b) edge[->] (t)
    ;
  }
  \caption{Qualifier hierarchy for effect type refinements.}
  \label{fig:effect_lattice}
\end{figure}

First, we refine the types of all methods and callback-like objects with an {\em effect qualifier} from Fig.~\ref{fig:effect_lattice}, which places an upper bound on the side effects they may incur, as in Gordon et al.~\cite{DBLP:conf/ecoop/GordonDEG13}.
We say a method is ``UI-effectful'' when it may access the UI and is annotated with \code{@UIEffect}.

\begin{figure}
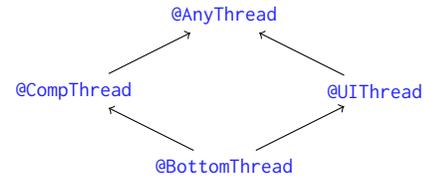

  \centering
  \tikz {
    \node (t) at (2,3) [] {\code{@AnyThread}};
    \node (l) at (0,2) [] {\code{@CompThread}};
    \node (r) at (4,2) [] {\code{@UIThread}};
    \node (b) at (2,1) [] {\code{@BottomThread}};
    \draw (b) edge[->] (l)
    (b) edge[->] (r)
    (l) edge[->] (t)
    (r) edge[->] (t)    
    ;
  }
  \caption{Qualifier hierarchy for thread type refinements.}
  \label{fig:thread_lattice}
\end{figure}

Next, we refine the types of all data streams with a {\em thread qualifier} from Fig.~\ref{fig:thread_lattice}, which places an upper bound on the thread on which they may emit events.
If such a thread cannot be determined statically, the stream will have the trivial \code{@AnyThread} qualifier, which does not restrict the base type at all.

In both cases, we take care to minimize annotation burden using sensible default qualifiers, shorthands to annotate entire classes or packages, and type inference to determine qualifiers where they are implied.
 
Finally, we typecheck the program, confirming that all assignments, return values, and function arguments respect their declared types, UI-effectful methods are never called from safe-effectful methods, and UI-effectful callbacks are subscribed only to UI-threaded streams.
Otherwise, we report warnings to the developer where these conditions are not met.

A more precise and formal treatment of the refinement type system we instantiate can be found in section~\ref{sec:semantics}.
For now, we will build intuition by applying our refinement typechecker to the program in Fig.~\ref{fig:motivating_example}.

When a developer tries to compile the program in Fig.~\ref{fig:motivating_example}, our compiler issues the following error at the \code{subscribe} callsite:

\begin{quote}
  \relsize{-0.5}\code{error: [rx.thread.violation] Subscribing a callback with @UIEffect to an observable scheduled on @CompThread; @UIEffect effects are limited to @UIThread observables}
\end{quote}

This error indicates a potential threading error in the program: the callbacks being passed to \code{subscribe} can touch the UI, but they are being subscribed to a \code{Observable} scheduled on a background computation thread.

The reason for this error is that \code{delay} automatically returns a stream running on a background computation thread.
Thus, even though the developer used \code{observeOn} to force the stream onto the main thread before subscribing UI-effectful callbacks, the intermediate call to \code{delay} promptly moved it to a computation thread.
In practice, we found that even expert developers were often unaware of this \code{delay} behavior, documentation of which is limited to a short note buried deep in a large generated documentation file.

\begin{sloppypar}
However, our typechecker refines the types of each program value: \code{@AnyThread} for the initial \code{carLocationData} stream, \code{@AnyThread} for the result of \code{filter}, \code{@UIThread} for the result of \code{observeOn}, \code{@CompThread} for the result of \code{delay}, and \code{@UIEffect} for the lambda expressions passed to \code{subscribe}.
Thus, our typechecker is able to identify the bug -- subscription of a \code{@UIEffect} function onto a \code{@CompThread} stream -- and issues the error reproduced above.
\end{sloppypar}

The fix to this issue, once the developer has been made aware of it by the typechecker, is simple and requires no annotations: swap the positions of the \code{delay} and \code{observeOn} calls and the code will typecheck and compile without error.

\vspace{\customSectionSpacing}\section{Thread \& Effect Semantics}\label{sec:semantics}

This section details the thread semantics of the Android UI framework and the ReactiveX threading model and formalizes the thread and effect type systems we use to analyze ReactiveX-based Android applications.

Recall that the Android UI toolkit is not thread-safe and adheres to a single thread model;
as such, any code that accesses UI components {\em must} do so from the UI thread~\cite{android_dev_guide}.
This is a relatively standard model for other UI frameworks, including iOS as well as Java's Swing, SWT, and AWT~\cite{ios}.

Existing work has established {\em effect types} as a useful abstraction for developers to avoid violating this single-thread assumption~\cite{DBLP:conf/ecoop/GordonDEG13}.

However, such work is limited to relatively simple threading models where the library provides an interface to the UI thread for code running elsewhere: for example, the Eclipse SWT UI framework defines a function with signature

\code{static void asyncExec(Runnable r);}

\hspace{-1em}which allows a developer to pass some UI-effectful code \code{r} to be run on the UI thread.

While Android does provide analogous functions (e.g., \code{Activity\#} \code{runOnUiThread}, \code{View\#post}) that can be analyzed by existing effect-typing techniques, it also provides more expressive functions beyond their reach (e.g., \code{AsyncTask}, \code{Handler}).
In addition, Reactive Extensions' threading constructs introduce even more complexity and require new techniques to be analyzed properly.

\vspace{\customSectionSpacing}\subsection{Effects}\label{sec:effect_semantics}
The application of effect type systems to UI frameworks is a fairly well-understood technique: functions are annotated with effect qualifiers, which can then be checked to verify that the annotated function does not perform any effects not permitted by its annotation.
Our effect system builds on top of that of Gordon et al.~\cite{DBLP:conf/ecoop/GordonDEG13}, wherein functions have one of two effect annotations: \code{@UIEffect}, denoting a method that may (or may not) interact with the UI, or \code{@SafeEffect}, denoting a method that is guaranteed not to touch the UI%
\footnote{Note that this ignores polymorphic qualifiers, which will be detailed in section~\ref{sec:qual_polymorphism}.}.

The sub-effecting relation $\preceq$ is given by \code{@SafeEffect} $\preceq$ \code{@UIEffect} and the two reflexive relationships, \code{@SafeEffect} $\preceq$ \code{@SafeEffect} and  \code{@UIEffect} $\preceq$ \code{@UIEffect}.
That is, a method with safe effect can be used in place of a method with UI effect, but not vice versa.

We say that a program is {\em effect-safe} when its effect annotations over-approximate all possible effects performed at runtime; thus, in an effect-safe program, a method annotated \code{@SafeEffect} is guaranteed never to interact with the UI.
Checking effect-safety of a program whose methods are annotated by their effect reduces to checking the following two conditions:
\begin{itemize}
\item[-] {\em Transitivity:} A method with effect annotation $e$ may only call a method with effect annotation $e'$ if $e'\preceq e$.
\item[-] {\em Inheritance:} A method with effect annotation $e$ may only override a method with effect annotation $e'$ if $e\preceq e'$.
\end{itemize}
In other words, the {\em transitivity} condition states that safe methods cannot call UI methods, while the {\em inheritance} condition states that methods cannot have UI effect if they override a safe method.
Assuming that UI-effectful library methods are annotated accordingly, proving these two conditions suffices to show that the UI is only accessed from methods annotated with \code{@UIEffect}.
Figure~\ref{fig:effect_examples} provides concrete examples of effect-safety violations.

\begin{figure}
  \begin{lstlisting}
      class A {
        @UIEffect   void foo() {...}
        @SafeEffect void bar() {...} }
      class B extends A {
        // Transitivity violation
        @SafeEffect void baz() { foo(); }
        // Inheritance violation
        @UIEffect void bar() {...} }
  \end{lstlisting}
  \caption{Example violations of the effect type system.  \code{B\#baz} violates the transitivity condition because it is annotated as safe but calls a UI-effectful method \code{A\#foo}, while \code{B\#bar} violates the inheritance condition because it manipulates the UI but overrides a method declared to be safe.}
  \label{fig:effect_examples}
\end{figure}

The reader may find it useful to think of these two conditions in terms of reachability in a directed graph whose vertices are methods, with edges from callers to callees and from superclass methods to overriding subclass methods.
In such a graph, an edge from a method with effect $e$ to a method with effect $e'$ violates one of the above conditions when $e\prec e'$.
The task of applying effect types to an Android application is thus equivalent to determining the region of nodes from which Android UI methods are reachable: that region has UI effect, while its complement has safe effect.

Other than the lambda support and inference mechanism described in section~\ref{sec:lambda_inference}, this effect type system is identical to that of Gordon et al. \cite{DBLP:conf/ecoop/GordonDEG13}.

\vspace{\customSectionSpacing}\subsection{Threads}
The threading behavior of Android applications that make use of stream-based programming frameworks like Reactive Extensions is determined not only by a small set of Android API methods with fixed semantics but also by a wide range of stream operators with dynamic threading behavior.
Effect typing alone is therefore insufficient to properly verify the UI thread-safety of such applications.

Consider, for example, the \code{subscribe} method of \code{Observable}, which is called on a stream in order to register some callback to be executed whenever an event is emitted by the receiver stream.

In contrast to Android's \code{runOnUiThread}, which can safely be passed a UI-effectful callback in all contexts, \code{subscribe} can only be passed a UI-effectful callback \code{obs} when the receiver stream is running on the UI thread.

In order to express that invariant, an analysis must reason not only about the effects of methods, but also the threads on which streams emit events (and, by extension, execute subscribed callbacks).

To that end, we augment our type system with type annotations that refine stream types by their thread.
These type annotations are drawn from the qualifier hierarchy given in Fig.~\ref{fig:thread_lattice}.
The top of the qualifier hierarchy, \code{@AnyThread}, denotes a stream that can emit events on any thread; \code{@UIThread} and \code{@CompThread} denote streams that can emit events only on the UI thread or a background computation thread, respectively; \code{@BottomThread} denotes a stream that cannot emit events on any thread.
This bottom type is never written by a programmer but is used within the typechecker for the \code{null} value, dead code, and wildcard lower bounds~\cite{cf_website}. 

Annotating stream operators with thread type refinements allows a typechecker to reason about the threading behavior of those constructs.
For example, the thread semantics of the \code{delay} function used in the motivating example in Fig.~\ref{fig:motivating_example} can be specified by annotating its receiver \code{@AnyThread} and its return type \code{@CompThread}.

Combining thread refinement types for streams with effect refinement types for methods is the key idea that allows our typechecker to verify UI thread-safety of stream-based Android applications by checking that subscription of UI-effectful callbacks only occurs on UI-threaded streams.

\vspace{\customSectionSpacing}\subsection{Qualifier Polymorphism}\label{sec:qual_polymorphism}
Some design patterns -- particularly those designed for modularity and reusability -- have effect and thread behavior that cannot be expressed by a single type signature with fixed refinements.
In these cases, we make use of {\em qualifier polymorphism}.

Qualifier polymorphism is a form of parametric polymorphism, which also underlies generics in Java and C\# and universally quantified types in Haskell and OCaml.
Consider, for example, this method from the Java collections library, which creates a singleton set:

\code{Set<T> singleton(T obj)\{...\}}

\hspace{-1em}The generic type variable \code{T} may be instantiated as any single Java object type, constraining the element type of the returned set to be the same as the argument type.

Similarly, qualifier polymorphism uses a generic refinement variable to {\em relate} type refinements rather than explicitly annotating types with a fixed qualifier.
We define a \code{@PolyThread} (resp., \code{@PolyUIEffect}) qualifier that may be instantiated as any concrete thread (resp., effect) qualifier, constraining the refinements on multiple types to be the same%
\footnote{
Implicitly, a class or method with polymorphic qualifier annotations is parameterized by a single refinement type variable which is used wherever the polymorphic qualifier is written.
As such, it is impossible to parameterize a definition by multiple refinement type variables, but we have not found any any code patterns in practice where such a type is required.
This polymorphic qualifier syntax is defined and provided by the Checker Framework~\cite{cf_website}.}.

Qualifier polymorphism is well suited to several design patterns in stream-based Android applications, several examples of which are selected and reproduced in Figure~\ref{fig:polymorphism}.

\begin{figure}
  \begin{lstlisting}
@PolyUIType interface Callback {
  @PolyUIEffect
  boolean handleMessage(Message m); }

class Observable<T> {
  @PolyThread Observable<T> take
   (@PolyThread Observable<T> this,
    int k){...};
  @PolyThread Observable<T> observeOn
   (@PolyThread Scheduler thread){...};}
  \end{lstlisting}
  \caption{Examples of thread- and effect-polymorphic types, drawn from Reactive Extensions' \code{io.reactivex.Observable} and Android's \code{android.os.Handler}, respectively.}
  \label{fig:polymorphism}
\end{figure}

The \code{Callback} interface exhibits effect polymorphism: we use the polymorphic qualifiers \code{@PolyUIType} and \code{@PolyUIEffect} to enforce that a \code{Callback} instance has a UI annotation when its \code{handleMessage} method has UI effect.
Similar interfaces such as \code{Runnable}, \code{Action}, and \code{Observer} are annotated analogously, relating the refinement type of the callback-like object to the effect of its implemented method(s).

Methods that take \code{Callback} instances with UI effect versions of \code{handleMessage} will need to declare the corresponding formal parameter as \code{@UI Callback}. Methods that do not care about the callback's effect take \code{@PolyUI Callback} instances. 
Our tool defaults to interpreting unannotated formals of a polymorphic type as \code{@AlwaysSafe} (non-UI affecting) instances. 
Analogous logic applies to the types of fields and locals.

The \code{Observable} class -- the main stream data type in ReactiveX -- exhibits two distinct forms of thread polymorphism.
First, most of its methods (e.g., \code{take} in Fig.~\ref{fig:polymorphism}) do not affect the thread of the stream they operate on; we express this behavior by constraining their receiver%
\footnote{\label{footnote_this_annotation}The receiver \code{this} is explicitly written and annotated as the first argument to the method, as per Java's Type Annotation specification;  this does not affect the semantics of the function whatsoever~\cite{jsr308}.}
and return values to have the same thread refinement with the \code{@PolyThread} qualifier.
Second, we use thread polymorphism to express the dependently-typed behavior of the \code{observeOn} operator, which takes a \code{Scheduler} (e.g., a thread pool) and returns a stream emitting events on that thread pool.
In order to do so, we overload the meaning of our thread qualifiers to apply to schedulers as well as streams and annotate the scheduler and the returned stream of \code{observeOn} with the \code{@PolyThread} qualifier.

Qualifier polymorphism introduces some additional complexity to the definition of effect-safety given in section~\ref{sec:effect_semantics}.
The interaction between polymorphic and concrete effects is fairly straightforward: we have  \code{@SafeEffect} $\preceq$ \code{@PolyUIEffect} and \code{@PolyUIEffect} $\preceq$ \code{@UIEffect}, since those relations would hold for whatever concrete effect \code{@PolyUIEffect} takes on.
That is, the body of an effect-polymorphic method can call any method with safe effect but is forbidden from calling UI-effectful methods.
On the other hand, the body of an effect-polymorphic method may only call other effect-polymorphic methods when they share the same receiver object, since arbitrary other effect-polymorphic methods may be instantiated with an incompatible concrete effect.

\vspace{\customSectionSpacing}\subsection{Lambdas and Qualifier Inference}\label{sec:lambda_inference}

The callbacks used in stream processing frameworks like ReactiveX are typically written as either anonymous inner classes or Java 8 lambda expressions.
When a callback is passed to a method whose formal parameter is annotated with a particular effect qualifier, the anonymous class or lambda must itself have a compatible type qualifier.
In the case of anonymous inner classes this can be achieved by using a type-use annotation (e.g., \code{new @UI Consumer\{...\}} to specify a \code{Consumer} whose \code{accept} method has \code{@UIEffect}). 
However, the syntax of Java lambda expressions does not permit any explicit type annotations (refinement or otherwise). 
Rather, their type is resolved through {\em type inference}.

Consider, for example, the Java type of the lambda expression passed as the first argument to \code{Observable\#subscribe} in Figure~\ref{fig:motivating_example}:

\code{car -> \{ /* display car on map */ \}}

The first formal parameter of \code{Observable\#subscribe} has type \code{@PolyUI Consumer}, so the compiler infers that \code{Consumer} is the {\em functional interface}%
\footnote{A functional interface is any Java interface with a single abstract method.} 
base type of the lambda.
During compilation, \code{javac} will convert any lambda expression to an anonymous instantiation of a compatible functional interface, inferring its base type from the context.

However, the base type inference mechanism does not apply to refinement type qualifiers.
Instead, we apply local type qualifier inference to compute the proper effect annotation with which to instantiate the \code{@PolyUI} polymorphic qualifier by inspecting the body of the lambda expression.
If a call to a method with \code{@UIEffect} effect is found within the body of the lambda, we mark the corresponding anonymous instance of the functional interface as \code{@UI}, and otherwise as safe.
In the example above, as long as the code in brackets includes at least one call to an \code{@UIEffect} method, we infer the type of the lambda to be \code{@UI Consumer}.

\vspace{\customSectionSpacing}\section{Typechecker Implementation}\label{sec:implementation}
\begin{figure}
  \centering
  \includegraphics[width=0.46\textwidth]{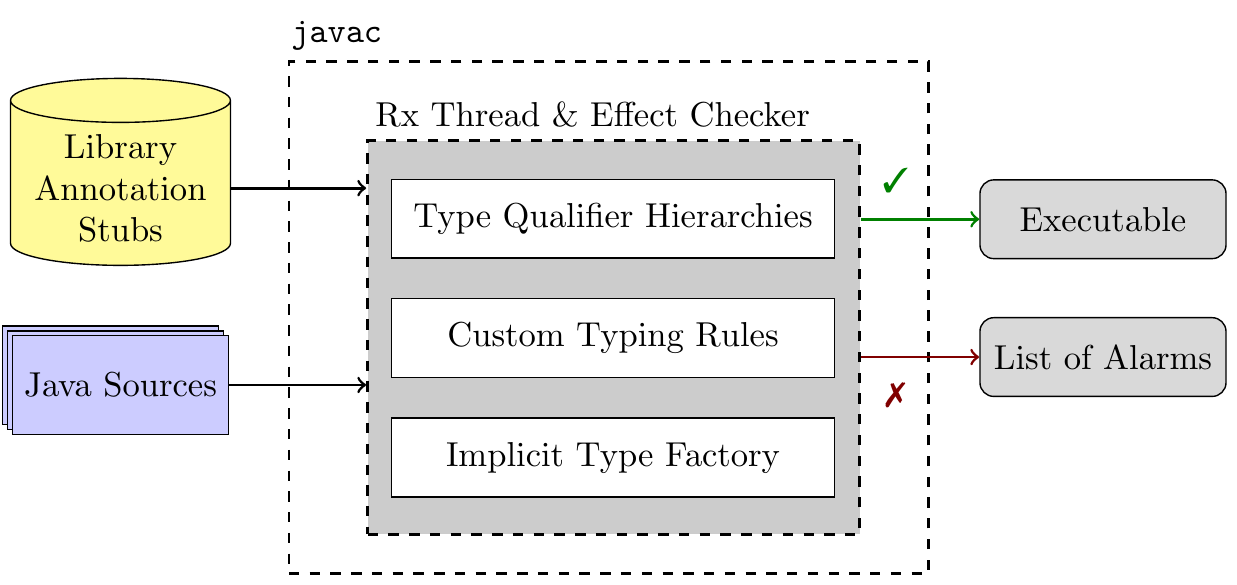}
  \caption{Typechecker infrastructure diagram showing its internal components, inputs and outputs.  From a user's perspective, UI thread-safety checking is integrated seamlessly with other \code{javac} compile-time checks in this default configuration, but it is also possible (by means of a command-line option) to run the checker as a standalone process, emitting a success message instead of generating an executable when the program is deemed safe.}
  \label{fig:system_diagram}
\end{figure}

This section details the implementation of a typechecker for the aforementioned thread and effect type systems, which is able to soundly verify the UI thread safety of ReactiveX-based Java Android applications.

The typechecker's design balances competing goals: it must not only provide reliable guarantees of safety and correctness, but also be easy-to-use and integrate into developer workflows.
What's more, in order to find and fix existing UI threading bugs, it must be applicable to legacy codebases without a prohibitive amount of configuration or annotation effort.
To that end, we build the typechecker upon the Checker Framework \cite{DBLP:conf/issta/PapiACPE08, cf_website, DBLP:conf/icse/DietlDEMS11}, which provides infrastructure for building custom typecheckers that augment Java's base type system.

Refinement types are implemented as Checker Framework type qualifiers: annotations on existing Java base types which are then processed at compile-time by our typechecker.
The Checker Framework allows users to check those types by simply invoking a \code{javac} compiler with certain command-line flags.
This design enables any modern Java build system or IDE to incorporate UI thread-safety typechecking with minimal configuration overhead.

The typechecker plugin itself consists primarily of three components, as shown in Fig.~\ref{fig:system_diagram}: the type qualifiers themselves, custom typing rules that specify the UI thread-safety invariant, and a type factory that generates annotations that are not explicitly written but rather inherited, derived, or inferred.

The {\em type qualifier hierarchies} -- described in detail individually in Section~\ref{sec:semantics} -- implement Java type annotations for each thread and effect type.
In addition to the type hierarchies shown in Figures~\ref{fig:effect_lattice} and~\ref{fig:thread_lattice}, this also includes polymorphic qualifiers that range over those lattices as well as class- and package-level annotations which apply some annotation to each method therein.

The {\em custom typing rules} express the constraints on effect inheritance and transitivity and enforce the core UI thread-safety invariant: that \code{@UIEffect}-ful callbacks may only be subscribed to \code{@UIThread} streams.
These custom typing rules augment the standard rules implemented by the Checker Framework which constrain, for example, actual parameters to declared formal parameter types, assignment {\em r}-values to corresponding {\em l}-value types, and return values to declared method signatures.

Finally, the {\em implicit type factory} is responsible for generating type refinements for those variables that lack an explicit refinement annotation.
This process takes one of three forms:
\begin{itemize}
\item[-] {\em Inheritance: }
  In addition to the \code{@UIEffect} and \code{@SafeEffect} method annotations, the effect type system we build upon also provides shorthand annotations to facilitate blanket annotation of all methods in a class or package~\cite{DBLP:conf/ecoop/GordonDEG13}.
  These shorthands are especially useful when all methods of a particular package or class share the same effect, saving the effort of annotating each one manually.
\item[-] {\em Inference: } 
  Whenever our typechecker encounters a lambda expression, we retrieve the functional interface base type inferred by \code{javac} for the lambda. 
  If that type has a concrete effect qualifier, no inference is required and we simply apply that qualifier.
  Otherwise, we scan the body of the lambda for any invocations of methods with \code{@UIEffect}.
  If any are found, then the anonymous instance associated with the lambda expression is annotated \code{@UI} (see Section \ref{sec:lambda_inference}).
\item[-] {\em Defaults: }
  When an un-annotated method or variable does not inherit a type refinement and inference does not apply, a sensible default is chosen by the typechecker.
  These defaults minimize annotation burden dramatically by only forcing developers to write annotations where the type refinement varies from the default.

  We refine un-annotated methods with the \code{@SafeEffect} qualifier so that annotations only need to be explicitly written on code that interacts with the UI, while un-annotated streams receive the \code{@AnyThread} trivial refinement so that, unless otherwise specified, we soundly assume a stream could emit events on any thread.

  The choice of default annotations is a design choice; in practice, we find that these defaults largely correspond to developer intuitions and reduce annotation overhead.
\end{itemize}

Those three elements -- qualifier hierarchies, custom typing rules, and implicit type generation -- suffice to typecheck {\em whole} programs: programs that are complete and self-contained in a set of source files.
The stream-based Android apps we are concerned with, however, are {\em open} programs which interact with other libraries and frameworks whose source code the typechecker may not have access to: ReactiveX and the Android standard library, at a bare minimum.

In order to typecheck code that interacts with un-analyzed external libraries, we refine types as needed at the public interface of libraries using {\em annotation stubs}: files consisting of annotated type signatures for library methods.
These annotations allow the typechecker to specify refinement types for third-party library code without checking the internals of those libraries or rebuilding them from source.

\begin{figure*}
  \begin{lstlisting}
    @UIType class ScrollView {//...
      @SafeEffect boolean post(@UI Runnable action);}
    class Observable<T> {//...
      @CompThread Observable<T> delay(long delay, TimeUnit unit);
      @PolyThread Observable<T> observeOn(@PolyThread Scheduler thread);
      @PolyThread Observable<T> take(@PolyThread Observable<T> this, int k);}
  \end{lstlisting}
  \vspace{-1em}
  \caption{Selected example stub annotations for ReactiveX Observable methods and an Android UI class.}
  \label{fig:example_stubs}
\end{figure*}

Annotation stubs are particularly effective for specifying the effects of Android library methods and the threading behavior of Observable operators.
We present several such annotation stubs in Figure~\ref{fig:example_stubs} in order to build intuition and demonstrate the technique.
\begin{itemize}
\item[-]
  \code{ScrollView} is an Android UI element, so most of its methods have UI effect.
  The \code{@UIType} class-level shorthand applies \code{@UIEffect} to each of its methods; however, some methods -- including \code{post} -- can safely be called from non-UI threads and are thus annotated \code{@SafeEffect}.
\item[-]
  \code{delay} schedules the returned \code{Observable} on the computation threadpool, so its return value is annotated \code{@CompThread}.
\item[-]
  \code{observeOn} returns an \code{Observable} that emits events on the given \code{Scheduler}.
  This behavior is expressed by applying the polymorphic \code{@PolyThread} qualifier to both the return type and the \code{thread} parameter, constraining them to both have the same thread refinement.
\item[-]
  \code{take} simply truncates a stream after \code{k} events and does not affect the thread of the stream it is called on, as is the case with most \code{Observable} operators.
  We therefore constrain its receiver%
  and return value to run on the same thread, using the polymorphic \code{@PolyThread} qualifier.
\end{itemize}

Annotation stubs are a possible source of unsoundness in our approach, since the refinement types they provide are trusted and the corresponding source code is unchecked.
However, we are able to mitigate this concern through careful review of source code and documentation for both Android and ReactiveX.
In particular, our annotation stubs conform to the Android developer guide's admonition that the \code{android.widget} and \code{android.view} packages comprise the Android UI toolkit (i.e. have UI effect) and to all threading behaviors specified in the ReactiveX \code{Observable} documentation~\cite{android_dev_guide,rx_website}.

\vspace{\customSectionSpacing}\section{Evaluation}
We answer the following research questions in order to evaluate our approach.
\begin{enumerate}
\item {\em Is the typechecker usable and practical?}
  Is the annotation burden sufficiently small for real-world Android developers to make use of the tool, and are the messages and warnings emitted by the tool useful and understandable?   
\item {\em Does the typechecker find real bugs and help fix them?}
  Are threading bugs in stream processing code prevalent in practice, does our tool identify them successfully, and is a program that successfully typechecks reliably free of such bugs?
\end{enumerate}
\vspace{\customSectionSpacing}\subsection{Evaluation Suite}
Our experimental evaluation consists of case studies over $8$ open-source Android applications, as well as a report on our experience applying the tool in production to two Android applications developed at \UBER

We select open-source applications for our experiments from Github according to the following criteria.
First, we restrict our search to Android applications written in Java (excluding Android applications written in other languages since our typechecker operates over a Java AST).

Of those, we consider only applications that import a 2.x version of ReactiveX.
We exclude applications using a 1.x version because our annotation stub coverage thereof is more sparse and applications using the older library version are more likely to be abandoned or broken.

Finally, we took the $8$ most recently indexed applications with 10 or more Github ``stars.''
These two criteria form an imperfect but practical proxy for repository activity: taking recently indexed apps avoids those that are abandoned or unmaintained, while requiring at least 10 stars ensures that the apps are not small personal projects or one-off experimental applications.

We believe that the $8$ subject programs selected thusly form a reasonably representative cross-section of open-source Android projects, including, for example:  a widely-forked template for Model-View-Presenter apps~\cite{benchmark6}, a client for a Russian technology news website~\cite{benchmark1}, and an app that scans, tracks, and organizes receipts~\cite{benchmark8}.
In total, this open-source evaluation suite consists of $142$ thousand lines of code written by $82$ distinct contributors.

In addition to this corpus of open-source applications, we have also applied the typechecker to the \UBEREATS~and \UBERDRIVER~apps, both of which are large closed-source Android applications written and maintained by professional developers at \UBER.
\vspace{\customSectionSpacing}\subsection{Annotation Workflow}
This section details the process of applying the typechecker to an existing Android application.

First, we clone the application, and confirm that it builds successfully in our local environment.  In practice, we found that all of the open-source apps gathered according to our criteria use the Gradle build system and were relatively easy to build locally~\cite{gradle}.
Next, we configure the application's build to invoke our custom typechecker simply by adding package dependencies and setting \code{javac} command-line options in the \code{build.gradle} configuration file.
At this point, the typechecker is fully configured and will be seamlessly integrated with the existing compilation process.

However, the task of refining effect and thread types of existing code remains.
We find it most efficient to annotate in two phases: first, apply effect annotations throughout the application, ignoring any errors related to threading and then, once all methods have the proper effect type, apply thread annotations and/or fix real bugs until no alarms remain.
 
Recall the two conditions that govern effect typing: methods may only call methods with lesser effect ({\em transitivity}) and override methods with greater effect ({\em inheritance}).

Since all unannotated application code has \code{@SafeEffect} by default and all Android UI methods have \code{@UIEffect} from annotation stubs, every call from the application into the Android UI toolkit violates the transitivity condition in the app's initial (unannotated) state.
In order to deal with this avalanche of alarms, we first identify all {\em packages} with names connoting UI effect (e.g., \code{ui}, \code{view}, \code{widget}, \code{layout}, etc.) and use the package-level \code{@UIPackage} annotation to apply \code{@UIEffect} to all methods therein.

Then, we triage remaining alarms and write type annotations manually until no alarms remain.
We proceed from files with the most alarms to those with fewest, using the class-level \code{@UIType} for especially alarm-ridden classes where most methods have UI effect.

It is possible that our package- and class-level annotations assign the \code{@UIEffect} refinement to methods that do not actually interact with the UI.
As such, manual triage is not as simple as blindly propagating \code{@UIEffect} annotations:
we are careful to identify cases where an alarm is due to the package- and class-level annotations and use \code{@SafeEffect} to exclude a method from the coarse UI annotation rather than further propagating the spurious \code{@UIEffect}.

Unlike the effect type system, which requires a moderate amount of manual annotation, the thread type system requires almost none.
Rather, our library annotation stubs are sufficient for the typechecker to determine the thread type refinement of intermediate values of fluent call-chains in most cases, and it is rare for developers to store \code{Observable} streams in instance fields or local variables, which would require type annotation at their declaration.

Therefore, once effect annotations are completed, the only remaining alarms in most applications are genuine threading violations, where a UI-effectful callback is subscribed to a non-UI-threaded stream.
We manually inspect these violations to confirm that they are bugs and then find fixes on a case-by-case basis.

Once fixes have been found and applied, the application is verifiably safe from non-UI thread access of UI elements in stream-based code.

Crucially, this fairly involved annotation process is a one-time cost.
After an application has been configured to use the thread and effect typechecker and fully annotated once, future bug fixes and feature additions require minimal annotations.
Nonetheless, possible threading errors are automatically reported by the compiler as they are written, allowing developers to write and modify complex asynchronous stream-based software with confidence.

\vspace{\customSectionSpacing}\subsection{Experimental Results}
\begin{figure*}[!ht]
  \centering
  \begin{tabular}{lrrrrr}
    App &
    KLoC &
    Annotations &
    Time Spent (hrs.) &
    Errors Found &
    Compile Time (sec.)\\
    \hline
    \code{ForPDA} \cite{benchmark1}&
    $33.0$& 
    $197$& 
    $3$& 
    $4$& 
    $27$\\ 
    \code{chat-sdk-android} \cite{benchmark2}&
    $34.6$& 
    $102$& 
    $2$& 
    $6$& 
    $21$\\ 
    \code{trust-wallet-android} \cite{benchmark3}&
    $8.8$& 
    $27$& 
    $1$& 
    $2$& 
    $17$\\ 
    \code{arch-components-date} \cite{benchmark4}&
    $0.7$& 
    $2$& 
    $0.5$& 
    $0$& 
    $8$\\ 
    \code{MVPArms} \cite{benchmark6}&
    $6.3$& 
    $59$& 
    $1$& 
    $1$& 
    $9$\\ 
    \code{rxbus} \cite{benchmark7}&
    $3.3$& 
    $12$& 
    $1$& 
    $0$& 
    $3$\\ 
    \code{SmartReceiptsLibrary} \cite{benchmark8}&
    $39.9$& 
    $217$& 
    $7$& 
    $16$& 
    $30$\\ 
    \code{OpenFoodFacts} \cite{benchmark9}&
    $14.9$& 
    $146$& 
    $3$& 
    $4$& 
    $41$\\ 
    \hline
    Averages&
    $17.7$&
    $95$&
    $2.3$&
    $4.1$&
    $19.5$
  \end{tabular}
  \caption{
    Open-source test corpus experimental results.
    Reported LoC figures are computed with \code{sloccount}~\cite{sloccount}.
    Reported compilation times are the mean of five executions of \code{gradle} compiling all application release sources with our typechecker enabled, measured on a laptop with an Intel i7-6700HQ processor and 8GB RAM, running Ubuntu 16.04.
  }
    \label{fig:experiment_results}
\end{figure*}

For our experiments, we apply the described workflow to the $8$ open-source applications gathered according to the aforementioned criteria.
Figure~\ref{fig:experiment_results} reports a summary of our results, which we apply here to the two stated research questions:

{\em Is the typechecker usable and practical?}

There are two facets to this question:
first, is the annotation burden reasonable, both in terms of developer time and total annotation, and
second, is the typechecker usable and understandable for real Android developers?

Our results on the open source evaluation suite demonstrate that, while the initial overhead of annotation is meaningful, it is far from prohibitive: we spent an average of $2.3$ hours per application, writing one annotation per $186$ source lines of code.


{\em Does the typechecker find real bugs and help fix them?}.

We find a total of $33$ threading defects spread across $6$ different apps in the open source evaluation suite, an average of $4.1$ errors per application.
The errors vary from simple oversights (e.g., forgetting an \code{observeOn}) to more complex interactions between multiple stream operators and combinators.
The wide range of bugs found across a majority of subject programs demonstrates the applicability and efficacy of our approach.

We reproduce one example defect identified by the typechecker here, taken from the \code{ForPDA}~\cite{benchmark1}~application and slightly modified for clarity.
\begin{quote}
\begin{lstlisting}
observable.onErrorReturn(throwable -> {
  handleError(throwable,onErrorAction);
  return fallbackValue; })
.observeOn(AndroidSchedulers.mainThread())
.subscribe(callback)
\end{lstlisting}
\end{quote}

The \code{onErrorReturn} method takes a lambda that runs whenever the receiver \code{Observable} emits an error, catching the error and emitting the lambda's return value instead.

In this case, the lambda simply calls \code{handleError} on the error \code{throwable} and returns a default value \code{fallbackValue}, before scheduling the resulting stream on the main thread and subscribing a callback \code{callback}.
The \code{handleError} function renders the error message in the UI, but it may run on a background thread since \code{onErrorReturn} precedes the \code{observeOn} that moves the stream to the main thread.
The fix to this bug is simple: switch the positions of \code{onErrorReturn} and \code{observeOn}, and the code compiles without error.
\vspace{\customSectionSpacing}\subsection{\UBER~Case Study}

As part of our evaluation, we deployed the typechecker in \UBER's development infrastructure for two major applications: the \UBEREATS~Android app and the newest version of the \UBERDRIVER~Android app (then in development, it was released to partners in April 2018).
The typechecker runs as part of the continuous integration pipeline in parallel with a manual code review process.
It inspects every patch, blocking code from merging into the trunk of the repository if it fails to typecheck.

Android apps at \UBER~are organized as a collection of targets, compiled and unit tested separately.
Over a period of $8$ months, we progressively enrolled individual targets from each of the two apps onto the typechecker.
For each target, we first wrote an initial set of annotations for existing code, then enabled typechecking for that target and merged in the new annotations simultaneously.

Overall, we enrolled over half a million LoC corresponding to targets from each of the two apps.
As of this writing, we did not enroll targets in other \UBER~apps, or shared platform code and first-party libraries, although we did add (trusted but unchecked) annotations stubs for shared UI code.

During the initial enrollment process, we made $41$ substantive changes to the application in addition to the added type annotations.
Most of those changes are simple additions of \code{observeOn} calls to ReactiveX fluent call chains to move streams onto the main thread.
These represent potential preexisting issues regarding improper UI access off the main thread. 
Although it is possible for many of them to be safe in practice, due to implicit knowledge about runtime behavior not captured in our type system, we consider the safety offered by the typechecker to be worth the potential marginal performance cost of moving these streams to the main thread or performing redundant \code{observeOn} calls.

During this enrollment period, we observed two critical production bugs involving ReactiveX streams accessing the UI from a background thread.
The first was distilled into the \code{delay()} example shown in Figure \ref{fig:motivating_example}.
The second was an issue where developers failed to notice that \code{Observable#switchMap} could create and return a new \code{Observable} on a different thread, rather than propagate the thread of its receiver.
These bugs were patched as soon as they manifested in production and the app updated accordingly,
but both would have been caught by the current version of the typechecker had it already been enabled for the corresponding targets.
This gives us some confidence that the tool is catching other similar issues before they make it past the development stage.
By design, issues identified by the typechecker are fixed before they become production bugs.

After enrollment, any further commits changing files inside the corresponding target are checked against our type system and developers are responsible for maintaining the annotations as part of the development cycle.
Over the $8$ months of our rollout, over $4,000$ commits and $178$ developers interacted with our typechecker. 
We note that enrolled targets contain an average of one type annotation per $104$ LoC, indicating that the burden placed on developers is relatively low.

Furthermore, when excluding commits made by the authors during the initial enrollment, the annotation burden falls to one type annotation per $179$ LoC, implying that much of the annotation burden is incurred upfront and not as ongoing maintenance cost.
 
We measured that developers have added $135$ \code{observeOn(...)} invocations to commits under code review in response to typechecker warnings, each representing a potential fix to a threading defect that could otherwise have gone uncaught.
Note that we cannot verify that all of these changes are fixing real threading bugs; some may have been needed due to a spurious \code{@UIEffect} annotation or other typechecking imprecision.
But, as we have not received developer feedback indicating excessive false positives, we believe that many of these changes were either fixing real issues or improving the code by making it more obviously safe.
Catching these threading issues early in the development process reduces costs and improves productivity, as developers do not need to context-switch back to code they wrote days or weeks earlier to fix bugs.

\vspace{\customSectionSpacing}\subsection{Threats to Validity}
The evaluation results presented in this section are predicated upon the correctness and soundness of our technique and experimental design.

First, it is important to note that our technique does not make any global guarantee about UI access or general thread-safety: the typechecker only verifies that UI-effectful code is annotated accordingly and stream-based code is safe from invalid UI thread access.
As such, our tool does {\em not} find UI threading bugs outside of stream-based code or non-UI threading bugs such as deadlocks or data races.

Furthermore, the safety guarantees provided by the typechecker are sound with respect to trusted library annotation stubs, which could potentially diverge from the true behavior of the underlying code.
We note, however, that our \code{@UIEffect} annotations for the Android standard library include all methods explicitly specified as UI-effectful by the official developer guide~\cite{android_dev_guide}.

Since these experiments were performed by an author of the tool, it is possible that the reported time spent underestimates the time that would be spent by a less experienced user.
However, that factor is balanced somewhat by the fact that we were unfamiliar with the open-source applications being annotated -- several of which were even documented in languages not spoken by the annotator.

It is also possible that our subject program corpus is not representative of stream-based Android programs in general, either due to small sample size or biases in our selection criteria.

\vspace{\customSectionSpacing}\section{Related Work}
Although there is a wealth of research on analysis of concurrent programs, existing approaches are ill-suited to the problem of detecting improperly threaded UI interactions in asynchronous stream-based software.

Much of the existing literature focuses on race detection, ordering violations, and deadlocks, leveraging a wide range of dynamic and static techniques including instrumentation, lockset algorithms, and model checking~\cite{DBLP:conf/pldi/NaikAW06,DBLP:conf/sosp/SavageABNS97,DBLP:conf/pldi/HenzingerJM04, DBLP:conf/pldi/FlanaganF09}.

However, the single thread model we wish to verify precludes the need for general thread analyses: there are no data races or deadlocks in a UI library that runs on only one thread.
Furthermore, the aforementioned analyses are computationally expensive, relying on various pointer and alias analyses to reason about Java's heap-allocated threading constructs.

Our type-based approach builds upon a deep body of work in extensible type systems to address both of these issues with concurrency analysis, since typechecking is an intraprocedural analysis that scales well and our type system is designed specifically to identify improper UI thread access.

The idea of augmenting an existing type system with more expressive domain-specific types was introduced by Freeman and Pfenning in their seminal work refining algebraic datatypes in Standard ML~\cite{DBLP:conf/pldi/FreemanP91}.
However, their contributions are primarily theoretical, providing only a barebones proof-of-concept implementation.

Extending a language's syntax with {\em annotations} to express refinements, as originally proposed for Standard ML in \cite{DBLP:conf/amast/Davies97}, has emerged as a technique enabling practical adoption of refinement type systems.
Annotation-based type systems are widespread,
augmenting an existing static type system in Haskell \cite{DBLP:conf/esop/VazouRJ13} and SML \cite{DBLP:conf/plpv/Dunfield07}
or adding simple static types to dynamically typed languages such as JavaScript \cite{typescript} and Python \cite{DBLP:conf/dls/VitousekKSB14}.

In Java, the Checker Framework \cite{DBLP:conf/issta/PapiACPE08} provides a means to extend the Java base type system with arbitrary custom type refinements.
This has also been used in the past, for example, to infer and check locking disciplines for multithreaded programs, verify information flow properties, and implement ownership and universe types~\cite{DBLP:conf/ccs/ErnstJMDPRKBBHVW14,DBLP:conf/icse/ErnstLMST16,DBLP:conf/ecoop/HuangDME12}.

The most closely related work to our own of which we are aware is the effect type system of Gordon et al.~\cite{DBLP:conf/ecoop/GordonDEG13}, which is also implemented using the Checker Framework.
However, without the qualifier inference and thread typing techniques detailed in this paper, their typechecker is unable to verify the stream-based applications with which we are concerned.

Testing is a popular alternative to static analysis (type-based or otherwise) when trying to rule out classes of errors in software.
There is a large body of work on automatically testing Android applications at the level of the UI.
Fully automated testing approaches include random and search-directed event generation tools \cite{Google:Monkey, DBLP:conf/sigsoft/MachiryTN13, DBLP:conf/issta/MaoHJ16, DBLP:conf/sigsoft/ClappBAA16, DBLP:conf/kbse/JiangWLC17, DBLP:conf/issta/SasnauskasR14, DBLP:conf/momm/YeCZJ13}, model-based exploration \cite{DBLP:conf/kbse/AmalfitanoFTCM12, DBLP:journals/software/AmalfitanoFTTM15, DBLP:conf/codaspy/RastogiCE13, DBLP:conf/fase/YangPX13, DBLP:conf/oopsla/ChoiNS13, DBLP:conf/oopsla/AzimN13, DBLP:conf/mobisys/Hao0NHG14, DBLP:conf/icse/MirzaeiGBSM16}, concolic testing \cite{DBLP:conf/sigsoft/AnandNHY12}, and event generation using evolutionary algorithms \cite{DBLP:conf/sigsoft/MahmoodMM14}.

Industrial state of the art mostly uses scripted UI tests constructed for a particular app using a testing framework \cite{Google:Espresso, Xamarin:Calabash, Robotium, Selendroid}.
Record-and-replay systems also see widespread use to isolate bugs encountered during manual testing or after deployment \cite{DBLP:conf/icse/GomezNAM13, DBLP:conf/oopsla/HuAN15, DBLP:conf/ispass/HalpernZPR15} and can be combined with search-based approaches into hybrid testing schemes~\cite{DBLP:conf/kbse/MaoHJ17}.

However, by their very nature, testing techniques cannot show the absence of bugs, only their presence.
Even a theoretical perfect tester which explores all possible UI interaction traces can miss threading bugs.
For example, issues flagged by our typechecker may depend on interaction with the network causing an event to be emitted on a stream that is subscribed by UI affecting code,
but it is possible to explore every reachable view without this event ever occurring and thus miss the bug. 

A survey by Choudhary et al. \cite{DBLP:conf/kbse/ChoudharyGO15} suggests that existing fully automated UI testing tools do not significantly outperform random exploration in terms of code coverage under a fixed time budget. 
This could be due to either a faster rate of event generation for the simpler approach or the difference in engineering efforts going into an industry standard tool versus research prototypes. 
In either case, these approaches are challenging to scale for large and complex applications.

Finally, testing-based tools usually run late in the development process, whereas our analysis can run before a code change has even been merged into the trunk of the repository for our apps.
\vspace{\customSectionSpacing}\section{Conclusion}
We present in this paper a technique that statically verifies UI access to occur only from the UI thread in stream-based Android programs.
Our approach refines the types of data streams with a static bound on their thread, refines the types of methods and callback-like objects with a static bound on their side-effects, and collates the two type systems in order to detect invalid thread accesses or prove the absence thereof.

We demonstrate the efficacy and usefulness of our typechecker by evaluating it on $8$ open-source applications and finding $33$ bugs.
We also report on our experience applying the tool at scale to two large production Android applications at \UBER, where it has analyzed over $4000$ commits by $178$ developers at time of writing.
\vspace{\customSectionSpacing}\section*{Acknowledgment}
The authors would like to thank Werner Dietl, Colin Gordon, Michael Ernst, and members of the CUPLV lab for valuable discussions.

This material is based on research sponsored in part by the National Science Foundation under grant number CCF-1055066 and by DARPA under agreement number FA8750-14-2-0263.
The U.S. Goverernment is authorized to reproduce and distribute reprints for Governmental purposes notwithstanding any copyright notation thereon.

\bibliographystyle{ACM-Reference-Format}
\bibliography{paper.bib} 


\begin{thebibliography}{64}


\ifx \showCODEN    \undefined \def \showCODEN     #1{\unskip}     \fi
\ifx \showDOI      \undefined \def \showDOI       #1{#1}\fi
\ifx \showISBNx    \undefined \def \showISBNx     #1{\unskip}     \fi
\ifx \showISBNxiii \undefined \def \showISBNxiii  #1{\unskip}     \fi
\ifx \showISSN     \undefined \def \showISSN      #1{\unskip}     \fi
\ifx \showLCCN     \undefined \def \showLCCN      #1{\unskip}     \fi
\ifx \shownote     \undefined \def \shownote      #1{#1}          \fi
\ifx \showarticletitle \undefined \def \showarticletitle #1{#1}   \fi
\ifx \showURL      \undefined \def \showURL       {\relax}        \fi
\providecommand\bibfield[2]{#2}
\providecommand\bibinfo[2]{#2}
\providecommand\natexlab[1]{#1}
\providecommand\showeprint[2][]{arXiv:#2}

\bibitem[\protect\citeauthoryear{Amalfitano, Fasolino, Tramontana, Carmine, and
  Memon}{Amalfitano et~al\mbox{.}}{2012}]%
        {DBLP:conf/kbse/AmalfitanoFTCM12}
\bibfield{author}{\bibinfo{person}{Domenico Amalfitano},
  \bibinfo{person}{Anna~Rita Fasolino}, \bibinfo{person}{Porfirio Tramontana},
  \bibinfo{person}{Salvatore~De Carmine}, {and} \bibinfo{person}{Atif~M.
  Memon}.} \bibinfo{year}{2012}\natexlab{}.
\newblock \showarticletitle{{Using {GUI} Ripping for Automated Testing of
  {Android} Applications}}. In \bibinfo{booktitle}{\emph{Proceedings of the
  {IEEE/ACM} International Conference on Automated Software Engineering
  ({ASE})}}. \bibinfo{pages}{258--261}.
\newblock


\bibitem[\protect\citeauthoryear{Amalfitano, Fasolino, Tramontana, Ta, and
  Memon}{Amalfitano et~al\mbox{.}}{2015}]%
        {DBLP:journals/software/AmalfitanoFTTM15}
\bibfield{author}{\bibinfo{person}{Domenico Amalfitano},
  \bibinfo{person}{Anna~Rita Fasolino}, \bibinfo{person}{Porfirio Tramontana},
  \bibinfo{person}{Bryan~Dzung Ta}, {and} \bibinfo{person}{Atif~M. Memon}.}
  \bibinfo{year}{2015}\natexlab{}.
\newblock \showarticletitle{{{MobiGUITAR}: Automated Model-Based Testing of
  Mobile Apps}}.
\newblock \bibinfo{journal}{\emph{{IEEE} Software}} \bibinfo{volume}{32},
  \bibinfo{number}{5} (\bibinfo{year}{2015}), \bibinfo{pages}{53--59}.
\newblock


\bibitem[\protect\citeauthoryear{Anand, Naik, Harrold, and Yang}{Anand
  et~al\mbox{.}}{2012}]%
        {DBLP:conf/sigsoft/AnandNHY12}
\bibfield{author}{\bibinfo{person}{Saswat Anand}, \bibinfo{person}{Mayur Naik},
  \bibinfo{person}{Mary~Jean Harrold}, {and} \bibinfo{person}{Hongseok Yang}.}
  \bibinfo{year}{2012}\natexlab{}.
\newblock \showarticletitle{{Automated Concolic Testing of Smartphone Apps}}.
  In \bibinfo{booktitle}{\emph{Proceedings of the {ACM} {SIGSOFT} International
  Symposium on Foundations of Software Engineering ({FSE})}}.
  \bibinfo{pages}{59}.
\newblock


\bibitem[\protect\citeauthoryear{Apple}{Apple}{2018a}]%
        {ios}
\bibfield{author}{\bibinfo{person}{Apple}.} \bibinfo{year}{2018}\natexlab{a}.
\newblock \bibinfo{booktitle}{\emph{{App Programming Guide for iOS}}}.
\newblock
\newblock
\shownote{\url{https://developer.apple.com/library/content/documentation/iPhoneOSProgrammingGuide}.}


\bibitem[\protect\citeauthoryear{Apple}{Apple}{2018b}]%
        {cocoa}
\bibfield{author}{\bibinfo{person}{Apple}.} \bibinfo{year}{2018}\natexlab{b}.
\newblock \bibinfo{booktitle}{\emph{{MacOS Cocoa}}}.
\newblock
\newblock
\shownote{\url{http://developer.apple.com/technologies/mac/cocoa.html}.}


\bibitem[\protect\citeauthoryear{Azim and Neamtiu}{Azim and Neamtiu}{2013}]%
        {DBLP:conf/oopsla/AzimN13}
\bibfield{author}{\bibinfo{person}{Tanzirul Azim} {and} \bibinfo{person}{Iulian
  Neamtiu}.} \bibinfo{year}{2013}\natexlab{}.
\newblock \showarticletitle{{Targeted and Depth-First Exploration for
  Systematic Testing of {Android} Apps}}. In
  \bibinfo{booktitle}{\emph{Proceedings of the {ACM} {SIGPLAN} International
  Conference on Object Oriented Programming Systems Languages {\&} Applications
  ({OOPSLA})}}. \bibinfo{pages}{641--660}.
\newblock


\bibitem[\protect\citeauthoryear{Baumann}{Baumann}{2018}]%
        {benchmark8}
\bibfield{author}{\bibinfo{person}{Will Baumann}.}
  \bibinfo{year}{2018}\natexlab{}.
\newblock \bibinfo{booktitle}{\emph{{Smart Receipts}}}.
\newblock
\newblock
\shownote{\url{https://github.com/wbaumann/SmartReceiptsLibrary}.}


\bibitem[\protect\citeauthoryear{{Chat SDK}}{{Chat SDK}}{2018}]%
        {benchmark2}
\bibfield{author}{\bibinfo{person}{{Chat SDK}}.}
  \bibinfo{year}{2018}\natexlab{}.
\newblock \bibinfo{booktitle}{\emph{{Chat SDK Android}}}.
\newblock
\newblock
\shownote{\url{https://github.com/chat-sdk/chat-sdk-android}.}


\bibitem[\protect\citeauthoryear{{Checker Framework Developers}}{{Checker
  Framework Developers}}{2018}]%
        {cf_website}
\bibfield{author}{\bibinfo{person}{{Checker Framework Developers}}.}
  \bibinfo{year}{2018}\natexlab{}.
\newblock \bibinfo{booktitle}{\emph{{Checker Framework Manual}}}.
\newblock
\newblock
\shownote{\url{https://checkerframework.org/manual}.}


\bibitem[\protect\citeauthoryear{Choi, Necula, and Sen}{Choi
  et~al\mbox{.}}{2013}]%
        {DBLP:conf/oopsla/ChoiNS13}
\bibfield{author}{\bibinfo{person}{Wontae Choi}, \bibinfo{person}{George~C.
  Necula}, {and} \bibinfo{person}{Koushik Sen}.}
  \bibinfo{year}{2013}\natexlab{}.
\newblock \showarticletitle{{Guided {GUI} Testing of {Android} Apps with
  Minimal Restart and Approximate Learning}}. In
  \bibinfo{booktitle}{\emph{Proceedings of the {ACM} {SIGPLAN} International
  Conference on Object Oriented Programming Systems Languages {\&} Applications
  ({OOPSLA})}}. \bibinfo{pages}{623--640}.
\newblock


\bibitem[\protect\citeauthoryear{Choudhary, Gorla, and Orso}{Choudhary
  et~al\mbox{.}}{2015}]%
        {DBLP:conf/kbse/ChoudharyGO15}
\bibfield{author}{\bibinfo{person}{Shauvik~Roy Choudhary},
  \bibinfo{person}{Alessandra Gorla}, {and} \bibinfo{person}{Alessandro Orso}.}
  \bibinfo{year}{2015}\natexlab{}.
\newblock \showarticletitle{{Automated Test Input Generation for {Android}: Are
  We There Yet?}}. In \bibinfo{booktitle}{\emph{Proceedings of the {IEEE/ACM}
  International Conference on Automated Software Engineering ({ASE})}}.
  \bibinfo{pages}{429--440}.
\newblock


\bibitem[\protect\citeauthoryear{Clapp, Bastani, Anand, and Aiken}{Clapp
  et~al\mbox{.}}{2016}]%
        {DBLP:conf/sigsoft/ClappBAA16}
\bibfield{author}{\bibinfo{person}{Lazaro Clapp}, \bibinfo{person}{Osbert
  Bastani}, \bibinfo{person}{Saswat Anand}, {and} \bibinfo{person}{Alex
  Aiken}.} \bibinfo{year}{2016}\natexlab{}.
\newblock \showarticletitle{{Minimizing {GUI} Event Traces}}. In
  \bibinfo{booktitle}{\emph{Proceedings of the {ACM} {SIGSOFT} International
  Symposium on Foundations of Software Engineering ({FSE})}}.
  \bibinfo{pages}{422--434}.
\newblock


\bibitem[\protect\citeauthoryear{Davies}{Davies}{1997}]%
        {DBLP:conf/amast/Davies97}
\bibfield{author}{\bibinfo{person}{Rowan Davies}.}
  \bibinfo{year}{1997}\natexlab{}.
\newblock \showarticletitle{{Refinement-Type Checker for Standard {ML}}}. In
  \bibinfo{booktitle}{\emph{Proceedings of the International Conference on
  Algebraic Methodology and Software Technology ({AMAST})}}.
  \bibinfo{pages}{565--566}.
\newblock


\bibitem[\protect\citeauthoryear{Dietl, Dietzel, Ernst, Muslu, and
  Schiller}{Dietl et~al\mbox{.}}{2011}]%
        {DBLP:conf/icse/DietlDEMS11}
\bibfield{author}{\bibinfo{person}{Werner Dietl}, \bibinfo{person}{Stephanie
  Dietzel}, \bibinfo{person}{Michael~D. Ernst}, \bibinfo{person}{Kivan{\c{c}}
  Muslu}, {and} \bibinfo{person}{Todd~W. Schiller}.}
  \bibinfo{year}{2011}\natexlab{}.
\newblock \showarticletitle{{Building and Using Pluggable Type-checkers}}. In
  \bibinfo{booktitle}{\emph{Proceedings of the International Conference on
  Software Engineering ({ICSE})}}. \bibinfo{pages}{681--690}.
\newblock


\bibitem[\protect\citeauthoryear{Dunfield}{Dunfield}{2007}]%
        {DBLP:conf/plpv/Dunfield07}
\bibfield{author}{\bibinfo{person}{Joshua Dunfield}.}
  \bibinfo{year}{2007}\natexlab{}.
\newblock \showarticletitle{{Refined Typechecking with Stardust}}. In
  \bibinfo{booktitle}{\emph{Proceedings of the {ACM} Workshop Programming
  Languages meets Program Verification ({PLPV})}}. \bibinfo{pages}{21--32}.
\newblock


\bibitem[\protect\citeauthoryear{Eclipse}{Eclipse}{2018}]%
        {eclipse}
\bibfield{author}{\bibinfo{person}{Eclipse}.} \bibinfo{year}{2018}\natexlab{}.
\newblock \bibinfo{booktitle}{\emph{{Standard Widget Toolkit}}}.
\newblock
\newblock
\shownote{\url{http://eclipse.org/swt}.}


\bibitem[\protect\citeauthoryear{Ernst, Just, Millstein, Dietl, Pernsteiner,
  Roesner, Koscher, Barros, Bhoraskar, Han, Vines, and Wu}{Ernst
  et~al\mbox{.}}{2014}]%
        {DBLP:conf/ccs/ErnstJMDPRKBBHVW14}
\bibfield{author}{\bibinfo{person}{Michael~D. Ernst},
  \bibinfo{person}{Ren{\'{e}} Just}, \bibinfo{person}{Suzanne Millstein},
  \bibinfo{person}{Werner Dietl}, \bibinfo{person}{Stuart Pernsteiner},
  \bibinfo{person}{Franziska Roesner}, \bibinfo{person}{Karl Koscher},
  \bibinfo{person}{Paulo Barros}, \bibinfo{person}{Ravi Bhoraskar},
  \bibinfo{person}{Seungyeop Han}, \bibinfo{person}{Paul Vines}, {and}
  \bibinfo{person}{Edward~XueJun Wu}.} \bibinfo{year}{2014}\natexlab{}.
\newblock \showarticletitle{{Collaborative Verification of Information Flow for
  a High-Assurance App Store}}. In \bibinfo{booktitle}{\emph{Proceedings of the
  {ACM} {SIGSAC} Conference on Computer and Communications Security ({CCS})}}.
  \bibinfo{pages}{1092--1104}.
\newblock


\bibitem[\protect\citeauthoryear{Ernst, Lovato, Macedonio, Spoto, and
  Thaine}{Ernst et~al\mbox{.}}{2016}]%
        {DBLP:conf/icse/ErnstLMST16}
\bibfield{author}{\bibinfo{person}{Michael~D. Ernst}, \bibinfo{person}{Alberto
  Lovato}, \bibinfo{person}{Damiano Macedonio}, \bibinfo{person}{Fausto Spoto},
  {and} \bibinfo{person}{Javier Thaine}.} \bibinfo{year}{2016}\natexlab{}.
\newblock \showarticletitle{{Locking Discipline Inference and Checking}}. In
  \bibinfo{booktitle}{\emph{Proceedings of the International Conference on
  Software Engineering ({ICSE})}}. \bibinfo{pages}{1133--1144}.
\newblock


\bibitem[\protect\citeauthoryear{Fazzini and Orso}{Fazzini and Orso}{2017}]%
        {DBLP:conf/kbse/FazziniO17}
\bibfield{author}{\bibinfo{person}{Mattia Fazzini} {and}
  \bibinfo{person}{Alessandro Orso}.} \bibinfo{year}{2017}\natexlab{}.
\newblock \showarticletitle{{Automated Cross-Platform Inconsistency Detection
  for Mobile Apps}}. In \bibinfo{booktitle}{\emph{Proceedings of the {IEEE/ACM}
  International Conference on Automated Software Engineering ({ASE}) 2017}}.
  \bibinfo{pages}{308--318}.
\newblock


\bibitem[\protect\citeauthoryear{Flanagan and Freund}{Flanagan and
  Freund}{2009}]%
        {DBLP:conf/pldi/FlanaganF09}
\bibfield{author}{\bibinfo{person}{Cormac Flanagan} {and}
  \bibinfo{person}{Stephen~N. Freund}.} \bibinfo{year}{2009}\natexlab{}.
\newblock \showarticletitle{{FastTrack: Efficient and Precise Dynamic Race
  Detection}}. In \bibinfo{booktitle}{\emph{Proceedings of the {ACM} {SIGPLAN}
  Conference on Programming Language Design and Implementation ({PLDI})}}.
\newblock


\bibitem[\protect\citeauthoryear{Franks}{Franks}{2018}]%
        {benchmark4}
\bibfield{author}{\bibinfo{person}{Rebecca Franks}.}
  \bibinfo{year}{2018}\natexlab{}.
\newblock \bibinfo{booktitle}{\emph{{Date Countdown}}}.
\newblock
\newblock
\shownote{\url{https://github.com/riggaroo/android-arch-components-date-countdown}.}


\bibitem[\protect\citeauthoryear{Freeman and Pfenning}{Freeman and
  Pfenning}{1991}]%
        {DBLP:conf/pldi/FreemanP91}
\bibfield{author}{\bibinfo{person}{Timothy~S. Freeman} {and}
  \bibinfo{person}{Frank Pfenning}.} \bibinfo{year}{1991}\natexlab{}.
\newblock \showarticletitle{{Refinement Types for ML}}. In
  \bibinfo{booktitle}{\emph{Proceedings of the {ACM} SIGPLAN Conference on
  Programming Language Design and Implementation (PLDI)}}.
  \bibinfo{pages}{268--277}.
\newblock


\bibitem[\protect\citeauthoryear{Gamma, Helm, Johnson, and Vlissides}{Gamma
  et~al\mbox{.}}{1995}]%
        {Gamma:1995:DPE:186897}
\bibfield{author}{\bibinfo{person}{Erich Gamma}, \bibinfo{person}{Richard
  Helm}, \bibinfo{person}{Ralph Johnson}, {and} \bibinfo{person}{John
  Vlissides}.} \bibinfo{year}{1995}\natexlab{}.
\newblock \bibinfo{booktitle}{\emph{Design Patterns: Elements of Reusable
  Object-oriented Software}}.
\newblock \bibinfo{publisher}{Addison-Wesley Longman Publishing Co., Inc.},
  \bibinfo{address}{Boston, MA, USA}.
\newblock
\showISBNx{0-201-63361-2}


\bibitem[\protect\citeauthoryear{Gomez, Neamtiu, Azim, and Millstein}{Gomez
  et~al\mbox{.}}{2013}]%
        {DBLP:conf/icse/GomezNAM13}
\bibfield{author}{\bibinfo{person}{Lorenzo Gomez}, \bibinfo{person}{Iulian
  Neamtiu}, \bibinfo{person}{Tanzirul Azim}, {and} \bibinfo{person}{Todd~D.
  Millstein}.} \bibinfo{year}{2013}\natexlab{}.
\newblock \showarticletitle{{{RERAN:} Timing- and Touch-Sensitive Record and
  Replay for {Android}}}. In \bibinfo{booktitle}{\emph{Proceedings of the
  International Conference on Software Engineering ({ICSE})}}.
  \bibinfo{pages}{72--81}.
\newblock


\bibitem[\protect\citeauthoryear{Google}{Google}{2018a}]%
        {android_dev_guide}
\bibfield{author}{\bibinfo{person}{Google}.} \bibinfo{year}{2018}\natexlab{a}.
\newblock \bibinfo{booktitle}{\emph{{Android Developer Guide}}}.
\newblock
\newblock
\shownote{\url{https://developer.android.com/guide}.}


\bibitem[\protect\citeauthoryear{Google}{Google}{2018b}]%
        {Google:Espresso}
\bibfield{author}{\bibinfo{person}{Google}.} \bibinfo{year}{2018}\natexlab{b}.
\newblock \bibinfo{booktitle}{\emph{{Espresso}}}.
\newblock
\newblock
\shownote{\url{https://developer.android.com/training/testing/ui-testing/espresso-testing.html}.}


\bibitem[\protect\citeauthoryear{Google}{Google}{2018c}]%
        {Google:Monkey}
\bibfield{author}{\bibinfo{person}{Google}.} \bibinfo{year}{2018}\natexlab{c}.
\newblock \bibinfo{booktitle}{\emph{{{UI/Application} Exerciser Monkey}}}.
\newblock
\newblock
\shownote{\url{https://developer.android.com/tools/help/monkey.html}.}


\bibitem[\protect\citeauthoryear{Gordon, Dietl, Ernst, and Grossman}{Gordon
  et~al\mbox{.}}{2013}]%
        {DBLP:conf/ecoop/GordonDEG13}
\bibfield{author}{\bibinfo{person}{Colin~S. Gordon}, \bibinfo{person}{Werner
  Dietl}, \bibinfo{person}{Michael~D. Ernst}, {and} \bibinfo{person}{Dan
  Grossman}.} \bibinfo{year}{2013}\natexlab{}.
\newblock \showarticletitle{{Java {UI} : Effects for Controlling {UI} Object
  Access}}. In \bibinfo{booktitle}{\emph{Proceedings of the European Conference
  on Object-Oriented Programming ({ECOOP})}}. \bibinfo{pages}{179--204}.
\newblock


\bibitem[\protect\citeauthoryear{{Gradle, Inc.}}{{Gradle, Inc.}}{2018}]%
        {gradle}
\bibfield{author}{\bibinfo{person}{{Gradle, Inc.}}}
  \bibinfo{year}{2018}\natexlab{}.
\newblock \bibinfo{booktitle}{\emph{Gradle Build Tool}}.
\newblock
\newblock
\shownote{\url{https://gradle.org}.}


\bibitem[\protect\citeauthoryear{Halpern, Zhu, Peri, and Reddi}{Halpern
  et~al\mbox{.}}{2015}]%
        {DBLP:conf/ispass/HalpernZPR15}
\bibfield{author}{\bibinfo{person}{Matthew Halpern}, \bibinfo{person}{Yuhao
  Zhu}, \bibinfo{person}{Ramesh Peri}, {and} \bibinfo{person}{Vijay~Janapa
  Reddi}.} \bibinfo{year}{2015}\natexlab{}.
\newblock \showarticletitle{{Mosaic: Cross-Platform User-Interaction Record and
  Replay for the Fragmented {Android} Ecosystem}}. In
  \bibinfo{booktitle}{\emph{Proceedings of the {IEEE} International Symposium
  on Performance Analysis of Systems and Software ({ISPASS})}}.
  \bibinfo{pages}{215--224}.
\newblock


\bibitem[\protect\citeauthoryear{Hao, Liu, Nath, Halfond, and Govindan}{Hao
  et~al\mbox{.}}{2014}]%
        {DBLP:conf/mobisys/Hao0NHG14}
\bibfield{author}{\bibinfo{person}{Shuai Hao}, \bibinfo{person}{Bin Liu},
  \bibinfo{person}{Suman Nath}, \bibinfo{person}{William G.~J. Halfond}, {and}
  \bibinfo{person}{Ramesh Govindan}.} \bibinfo{year}{2014}\natexlab{}.
\newblock \showarticletitle{{{PUMA:} Programmable UI-Automation for Large-scale
  Dynamic Analysis of Mobile Apps}}. In \bibinfo{booktitle}{\emph{Proceedings
  of the International Conference on Mobile Systems, Applications, and Services
  ({MobiSys})}}. \bibinfo{pages}{204--217}.
\newblock


\bibitem[\protect\citeauthoryear{Henzinger, Jhala, and Majumdar}{Henzinger
  et~al\mbox{.}}{2004}]%
        {DBLP:conf/pldi/HenzingerJM04}
\bibfield{author}{\bibinfo{person}{Thomas~A. Henzinger},
  \bibinfo{person}{Ranjit Jhala}, {and} \bibinfo{person}{Rupak Majumdar}.}
  \bibinfo{year}{2004}\natexlab{}.
\newblock \showarticletitle{{Race Checking by Context Inference}}. In
  \bibinfo{booktitle}{\emph{Proceedings of the {ACM} {SIGPLAN} Conference on
  Programming Language Design and Implementation ({PLDI})}}.
  \bibinfo{pages}{1--13}.
\newblock


\bibitem[\protect\citeauthoryear{Hu, Azim, and Neamtiu}{Hu
  et~al\mbox{.}}{2015}]%
        {DBLP:conf/oopsla/HuAN15}
\bibfield{author}{\bibinfo{person}{Yongjian Hu}, \bibinfo{person}{Tanzirul
  Azim}, {and} \bibinfo{person}{Iulian Neamtiu}.}
  \bibinfo{year}{2015}\natexlab{}.
\newblock \showarticletitle{{Versatile yet Lightweight Record-and-Replay for
  {Android}}}. In \bibinfo{booktitle}{\emph{Proceedings of the {ACM} {SIGPLAN}
  International Conference on Object-Oriented Programming, Systems, Languages,
  and Applications ({OOPSLA})}}. \bibinfo{pages}{349--366}.
\newblock


\bibitem[\protect\citeauthoryear{Huang, Dietl, Milanova, and Ernst}{Huang
  et~al\mbox{.}}{2012}]%
        {DBLP:conf/ecoop/HuangDME12}
\bibfield{author}{\bibinfo{person}{Wei Huang}, \bibinfo{person}{Werner Dietl},
  \bibinfo{person}{Ana Milanova}, {and} \bibinfo{person}{Michael~D. Ernst}.}
  \bibinfo{year}{2012}\natexlab{}.
\newblock \showarticletitle{{Inference and Checking of Object Ownership}}. In
  \bibinfo{booktitle}{\emph{Proceedings of the European Conference on
  Object-Oriented Programming ({ECOOP})}}. \bibinfo{pages}{181--206}.
\newblock


\bibitem[\protect\citeauthoryear{Jiang, Wu, Li, and Chan}{Jiang
  et~al\mbox{.}}{2017}]%
        {DBLP:conf/kbse/JiangWLC17}
\bibfield{author}{\bibinfo{person}{Bo Jiang}, \bibinfo{person}{Yuxuan Wu},
  \bibinfo{person}{Teng Li}, {and} \bibinfo{person}{W.~K. Chan}.}
  \bibinfo{year}{2017}\natexlab{}.
\newblock \showarticletitle{{SimplyDroid: Efficient Event Sequence
  Simplification for Android Application}}. In
  \bibinfo{booktitle}{\emph{Proceedings of the {IEEE/ACM} International
  Conference on Automated Software Engineering ({ASE})}}.
  \bibinfo{pages}{297--307}.
\newblock


\bibitem[\protect\citeauthoryear{{JSR 308 Expert Group}}{{JSR 308 Expert
  Group}}{2014}]%
        {jsr308}
\bibfield{author}{\bibinfo{person}{{JSR 308 Expert Group}}.}
  \bibinfo{year}{2014}\natexlab{}.
\newblock \bibinfo{booktitle}{\emph{{Annotations on Java Types}}}.
\newblock
\newblock
\shownote{\url{http://download.oracle.com/otndocs/jcp/annotations-2014_01_08-pfd-spec}.}


\bibitem[\protect\citeauthoryear{Kuwork}{Kuwork}{2018}]%
        {benchmark7}
\bibfield{author}{\bibinfo{person}{Kuwork}.} \bibinfo{year}{2018}\natexlab{}.
\newblock \bibinfo{booktitle}{\emph{{RxBus}}}.
\newblock
\newblock
\shownote{\url{https://github.com/kuwork/rxbus}.}


\bibitem[\protect\citeauthoryear{Machiry, Tahiliani, and Naik}{Machiry
  et~al\mbox{.}}{2013}]%
        {DBLP:conf/sigsoft/MachiryTN13}
\bibfield{author}{\bibinfo{person}{Aravind Machiry}, \bibinfo{person}{Rohan
  Tahiliani}, {and} \bibinfo{person}{Mayur Naik}.}
  \bibinfo{year}{2013}\natexlab{}.
\newblock \showarticletitle{{Dynodroid: An Input Generation System for
  {Android} Apps}}. In \bibinfo{booktitle}{\emph{Proceedings of the Joint
  Meeting on Foundations of Software Engineering ({FSE/ESEC})}}.
  \bibinfo{pages}{224--234}.
\newblock


\bibitem[\protect\citeauthoryear{Mahmood, Mirzaei, and Malek}{Mahmood
  et~al\mbox{.}}{2014}]%
        {DBLP:conf/sigsoft/MahmoodMM14}
\bibfield{author}{\bibinfo{person}{Riyadh Mahmood}, \bibinfo{person}{Nariman
  Mirzaei}, {and} \bibinfo{person}{Sam Malek}.}
  \bibinfo{year}{2014}\natexlab{}.
\newblock \showarticletitle{{{EvoDroid}: Segmented Evolutionary Testing of
  {Android} Apps}}. In \bibinfo{booktitle}{\emph{Proceedings of the {ACM}
  {SIGSOFT} International Symposium on Foundations of Software Engineering
  ({FSE})}}. \bibinfo{pages}{599--609}.
\newblock


\bibitem[\protect\citeauthoryear{Mao, Harman, and Jia}{Mao
  et~al\mbox{.}}{2016}]%
        {DBLP:conf/issta/MaoHJ16}
\bibfield{author}{\bibinfo{person}{Ke Mao}, \bibinfo{person}{Mark Harman},
  {and} \bibinfo{person}{Yue Jia}.} \bibinfo{year}{2016}\natexlab{}.
\newblock \showarticletitle{{Sapienz: Multi-Objective Automated Testing for
  Android Applications}}. In \bibinfo{booktitle}{\emph{Proceedings of the
  International Symposium on Software Testing and Analysis ({ISSTA})}}.
  \bibinfo{pages}{94--105}.
\newblock


\bibitem[\protect\citeauthoryear{Mao, Harman, and Jia}{Mao
  et~al\mbox{.}}{2017}]%
        {DBLP:conf/kbse/MaoHJ17}
\bibfield{author}{\bibinfo{person}{Ke Mao}, \bibinfo{person}{Mark Harman},
  {and} \bibinfo{person}{Yue Jia}.} \bibinfo{year}{2017}\natexlab{}.
\newblock \showarticletitle{{Crowd Intelligence Enhances Automated Mobile
  Testing}}. In \bibinfo{booktitle}{\emph{Proceedings of the {IEEE/ACM}
  International Conference on Automated Software Engineering ({ASE})}}.
  \bibinfo{pages}{16--26}.
\newblock


\bibitem[\protect\citeauthoryear{Microsoft}{Microsoft}{2018}]%
        {typescript}
\bibfield{author}{\bibinfo{person}{Microsoft}.}
  \bibinfo{year}{2018}\natexlab{}.
\newblock \bibinfo{booktitle}{\emph{{TypeScript}}}.
\newblock
\newblock
\shownote{\url{https://www.typescriptlang.org}.}


\bibitem[\protect\citeauthoryear{Mirzaei, Garcia, Bagheri, Sadeghi, and
  Malek}{Mirzaei et~al\mbox{.}}{2016}]%
        {DBLP:conf/icse/MirzaeiGBSM16}
\bibfield{author}{\bibinfo{person}{Nariman Mirzaei}, \bibinfo{person}{Joshua
  Garcia}, \bibinfo{person}{Hamid Bagheri}, \bibinfo{person}{Alireza Sadeghi},
  {and} \bibinfo{person}{Sam Malek}.} \bibinfo{year}{2016}\natexlab{}.
\newblock \showarticletitle{{Reducing Combinatorics in {GUI} Testing of Android
  Applications}}. In \bibinfo{booktitle}{\emph{Proceedings of the International
  Conference on Software Engineering {(ICSE)}}}. \bibinfo{pages}{559--570}.
\newblock


\bibitem[\protect\citeauthoryear{Naik, Aiken, and Whaley}{Naik
  et~al\mbox{.}}{2006}]%
        {DBLP:conf/pldi/NaikAW06}
\bibfield{author}{\bibinfo{person}{Mayur Naik}, \bibinfo{person}{Alex Aiken},
  {and} \bibinfo{person}{John Whaley}.} \bibinfo{year}{2006}\natexlab{}.
\newblock \showarticletitle{{Effective Static Race Detection for Java}}. In
  \bibinfo{booktitle}{\emph{Proceedings of the {ACM} {SIGPLAN} Conference on
  Programming Language Design and Implementation ({PLDI})}}.
  \bibinfo{pages}{308--319}.
\newblock


\bibitem[\protect\citeauthoryear{Nizamiev}{Nizamiev}{2018}]%
        {benchmark1}
\bibfield{author}{\bibinfo{person}{Evgeny Nizamiev}.}
  \bibinfo{year}{2018}\natexlab{}.
\newblock \bibinfo{booktitle}{\emph{{ForPDA}}}.
\newblock
\newblock
\shownote{\url{https://github.com/RadiationX/ForPDA}.}


\bibitem[\protect\citeauthoryear{{Open Food Facts, Org.}}{{Open Food Facts,
  Org.}}{2018}]%
        {benchmark9}
\bibfield{author}{\bibinfo{person}{{Open Food Facts, Org.}}}
  \bibinfo{year}{2018}\natexlab{}.
\newblock \bibinfo{booktitle}{\emph{{Open Food Facts}}}.
\newblock
\newblock
\shownote{\url{https://github.com/openfoodfacts/openfoodfacts-androidapp}.}


\bibitem[\protect\citeauthoryear{Oracle}{Oracle}{2018}]%
        {swing}
\bibfield{author}{\bibinfo{person}{Oracle}.} \bibinfo{year}{2018}\natexlab{}.
\newblock \bibinfo{booktitle}{\emph{{JDK Swing Framework}}}.
\newblock
\newblock
\shownote{\url{http://docs.oracle.com/javase/6/docs/technotes/guides/swing/}.}


\bibitem[\protect\citeauthoryear{Papi, Ali, Jr., Perkins, and Ernst}{Papi
  et~al\mbox{.}}{2008}]%
        {DBLP:conf/issta/PapiACPE08}
\bibfield{author}{\bibinfo{person}{Matthew~M. Papi}, \bibinfo{person}{Mahmood
  Ali}, \bibinfo{person}{Telmo Luis~Correa Jr.}, \bibinfo{person}{Jeff~H.
  Perkins}, {and} \bibinfo{person}{Michael~D. Ernst}.}
  \bibinfo{year}{2008}\natexlab{}.
\newblock \showarticletitle{{Practical Pluggable Types for Java}}. In
  \bibinfo{booktitle}{\emph{Proceedings of the {ACM/SIGSOFT} International
  Symposium on Software Testing and Analysis ({ISSTA})}}.
  \bibinfo{pages}{201--212}.
\newblock


\bibitem[\protect\citeauthoryear{Rastogi, Chen, and Enck}{Rastogi
  et~al\mbox{.}}{2013}]%
        {DBLP:conf/codaspy/RastogiCE13}
\bibfield{author}{\bibinfo{person}{Vaibhav Rastogi}, \bibinfo{person}{Yan
  Chen}, {and} \bibinfo{person}{William Enck}.}
  \bibinfo{year}{2013}\natexlab{}.
\newblock \showarticletitle{{{AppsPlayground}: Automatic Security Analysis of
  Smartphone Applications}}. In \bibinfo{booktitle}{\emph{Proceedings of the
  {ACM} Conference on Data and Application Security and Privacy {(CODASPY)}}}.
  \bibinfo{pages}{209--220}.
\newblock


\bibitem[\protect\citeauthoryear{{Reactive Extensions}}{{Reactive
  Extensions}}{2018}]%
        {rx_website}
\bibfield{author}{\bibinfo{person}{{Reactive Extensions}}.}
  \bibinfo{year}{2018}\natexlab{}.
\newblock \bibinfo{booktitle}{\emph{{Reactive Extensions}}}.
\newblock
\newblock
\shownote{\url{reactivex.io}.}


\bibitem[\protect\citeauthoryear{Robotium}{Robotium}{2018}]%
        {Robotium}
\bibfield{author}{\bibinfo{person}{Robotium}.} \bibinfo{year}{2018}\natexlab{}.
\newblock \bibinfo{booktitle}{\emph{{Robotium}}}.
\newblock
\newblock
\shownote{\url{https://github.com/robotiumtech/robotium}.}


\bibitem[\protect\citeauthoryear{Sasnauskas and Regehr}{Sasnauskas and
  Regehr}{2014}]%
        {DBLP:conf/issta/SasnauskasR14}
\bibfield{author}{\bibinfo{person}{Raimondas Sasnauskas} {and}
  \bibinfo{person}{John Regehr}.} \bibinfo{year}{2014}\natexlab{}.
\newblock \showarticletitle{{Intent Fuzzer: Crafting Intents of Death}}. In
  \bibinfo{booktitle}{\emph{Proceedings of the Joint International Workshop on
  Dynamic Analysis {(WODA)} and Software and System Performance Testing,
  Debugging, and Analytics {(PERTEA)}}}. \bibinfo{pages}{1--5}.
\newblock


\bibitem[\protect\citeauthoryear{Savage, Burrows, Nelson, Sobalvarro, and
  Anderson}{Savage et~al\mbox{.}}{1997}]%
        {DBLP:conf/sosp/SavageABNS97}
\bibfield{author}{\bibinfo{person}{Stefan Savage}, \bibinfo{person}{Michael
  Burrows}, \bibinfo{person}{Greg Nelson}, \bibinfo{person}{Patrick
  Sobalvarro}, {and} \bibinfo{person}{Thomas~E. Anderson}.}
  \bibinfo{year}{1997}\natexlab{}.
\newblock \showarticletitle{{Eraser: {A} Dynamic Data Race Detector for
  Multi-Threaded Programs}}. In \bibinfo{booktitle}{\emph{Proceedings of the
  {ACM} Symposium on Operating System Principles ({SOSP})}}.
  \bibinfo{pages}{27--37}.
\newblock


\bibitem[\protect\citeauthoryear{Selendroid}{Selendroid}{2018}]%
        {Selendroid}
\bibfield{author}{\bibinfo{person}{Selendroid}.}
  \bibinfo{year}{2018}\natexlab{}.
\newblock \bibinfo{booktitle}{\emph{{Selendroid}}}.
\newblock
\newblock
\shownote{\url{http://selendroid.io/}.}


\bibitem[\protect\citeauthoryear{Trust}{Trust}{2018}]%
        {benchmark3}
\bibfield{author}{\bibinfo{person}{Trust}.} \bibinfo{year}{2018}\natexlab{}.
\newblock \bibinfo{booktitle}{\emph{{Trust Wallet}}}.
\newblock
\newblock
\shownote{\url{https://github.com/TrustWallet/trust-wallet-android-source}.}


\bibitem[\protect\citeauthoryear{Vazou, Rondon, and Jhala}{Vazou
  et~al\mbox{.}}{2013}]%
        {DBLP:conf/esop/VazouRJ13}
\bibfield{author}{\bibinfo{person}{Niki Vazou}, \bibinfo{person}{Patrick~Maxim
  Rondon}, {and} \bibinfo{person}{Ranjit Jhala}.}
  \bibinfo{year}{2013}\natexlab{}.
\newblock \showarticletitle{{Abstract Refinement Types}}. In
  \bibinfo{booktitle}{\emph{Proceedings of the European Symposium on
  Programming ({ESOP})}}. \bibinfo{pages}{209--228}.
\newblock


\bibitem[\protect\citeauthoryear{Vitousek, Kent, Siek, and Baker}{Vitousek
  et~al\mbox{.}}{2014}]%
        {DBLP:conf/dls/VitousekKSB14}
\bibfield{author}{\bibinfo{person}{Michael~M. Vitousek},
  \bibinfo{person}{Andrew~M. Kent}, \bibinfo{person}{Jeremy~G. Siek}, {and}
  \bibinfo{person}{Jim Baker}.} \bibinfo{year}{2014}\natexlab{}.
\newblock \showarticletitle{{Design and Evaluation of Gradual Typing for
  Python}}. In \bibinfo{booktitle}{\emph{Proceedings of the {ACM} Symposium on
  Dynamic Languages ({DLS})}}. \bibinfo{pages}{45--56}.
\newblock


\bibitem[\protect\citeauthoryear{Wei, Liu, and Cheung}{Wei
  et~al\mbox{.}}{2016}]%
        {DBLP:conf/kbse/WeiLC16}
\bibfield{author}{\bibinfo{person}{Lili Wei}, \bibinfo{person}{Yepang Liu},
  {and} \bibinfo{person}{Shing{-}Chi Cheung}.} \bibinfo{year}{2016}\natexlab{}.
\newblock \showarticletitle{{Taming Android Fragmentation: Characterizing and
  Detecting Compatibility Issues for Android Apps}}. In
  \bibinfo{booktitle}{\emph{Proceedings of the {IEEE/ACM} International
  Conference on Automated Software Engineering ({ASE})}}.
  \bibinfo{pages}{226--237}.
\newblock


\bibitem[\protect\citeauthoryear{Wheeler}{Wheeler}{2004}]%
        {sloccount}
\bibfield{author}{\bibinfo{person}{David~A. Wheeler}.}
  \bibinfo{year}{2004}\natexlab{}.
\newblock \bibinfo{booktitle}{\emph{{SLOCCount}}}.
\newblock
\newblock
\shownote{\url{https://www.dwheeler.com/sloccount/}.}


\bibitem[\protect\citeauthoryear{Xamarin}{Xamarin}{2018}]%
        {Xamarin:Calabash}
\bibfield{author}{\bibinfo{person}{Xamarin}.} \bibinfo{year}{2018}\natexlab{}.
\newblock \bibinfo{booktitle}{\emph{{Calabash}}}.
\newblock
\newblock
\shownote{\url{http://calaba.sh/}.}


\bibitem[\protect\citeauthoryear{Yan}{Yan}{2018}]%
        {benchmark6}
\bibfield{author}{\bibinfo{person}{Jess Yan}.} \bibinfo{year}{2018}\natexlab{}.
\newblock \bibinfo{booktitle}{\emph{{MVP Arms}}}.
\newblock
\newblock
\shownote{\url{https://github.com/JessYanCoding/MVPArms}.}


\bibitem[\protect\citeauthoryear{Yang, Prasad, and Xie}{Yang
  et~al\mbox{.}}{2013}]%
        {DBLP:conf/fase/YangPX13}
\bibfield{author}{\bibinfo{person}{Wei Yang}, \bibinfo{person}{Mukul~R.
  Prasad}, {and} \bibinfo{person}{Tao Xie}.} \bibinfo{year}{2013}\natexlab{}.
\newblock \showarticletitle{{A Grey-Box Approach for Automated GUI-Model
  Generation of Mobile Applications}}. In \bibinfo{booktitle}{\emph{Proceedings
  of the International Conference on Fundamental Approaches to Software
  Engineering {(FASE)}}}. \bibinfo{pages}{250--265}.
\newblock


\bibitem[\protect\citeauthoryear{Ye, Cheng, Zhang, and Jiang}{Ye
  et~al\mbox{.}}{2013}]%
        {DBLP:conf/momm/YeCZJ13}
\bibfield{author}{\bibinfo{person}{Hui Ye}, \bibinfo{person}{Shaoyin Cheng},
  \bibinfo{person}{Lanbo Zhang}, {and} \bibinfo{person}{Fan Jiang}.}
  \bibinfo{year}{2013}\natexlab{}.
\newblock \showarticletitle{{DroidFuzzer: Fuzzing the {Android} Apps with
  Intent-Filter Tag}}. In \bibinfo{booktitle}{\emph{Proceedings of the
  International Conference on Advances in Mobile Computing {\&} Multimedia
  ({MoMM})}}. \bibinfo{pages}{68}.
\newblock


\bibitem[\protect\citeauthoryear{Zhang, L{\"{u}}, and Ernst}{Zhang
  et~al\mbox{.}}{2012}]%
        {DBLP:conf/issta/ZhangLE12}
\bibfield{author}{\bibinfo{person}{Sai Zhang}, \bibinfo{person}{Hao L{\"{u}}},
  {and} \bibinfo{person}{Michael~D. Ernst}.} \bibinfo{year}{2012}\natexlab{}.
\newblock \showarticletitle{{Finding Errors in Multithreaded {GUI}
  Applications}}. In \bibinfo{booktitle}{\emph{Proceedings of the {ACM/SIGSOFT}
  International Symposium on Software Testing and Analysis ({ISSTA})}}.
  \bibinfo{pages}{243--253}.
\newblock


\end{thebibliography}
\end{document}